\pdfoutput=1

\documentclass[preprint,showkeys,secnumarabic,amsfonts,showpacs,amsmath,amssymb]{revtex4}
\usepackage[dvips]{color}
\usepackage{epsfig}
\usepackage{amsmath}
\usepackage{graphicx}

\def\Box{\hbox{$\rlap{$\sqcup$}\sqcap$}}

\begin{document}

\title{\bf Stability analysis and Observational Measurement in Chameleonic Generalised Brans--Dicke Cosmology }
\author{Hossein Farajollahi}
\email{hosseinf@guilan.ac.ir}

\author{Amin Salehi}
\email{a.salehi@guilan.ac.ir}
\affiliation{Department of Physics, University of Guilan, Rasht, Iran}

\date{\today}

\begin{abstract}
We investigate the dynamics of the chameleonic Generalised Brans--Dicke model
in flat FRW cosmology. In a new approach, a framework to study
stability and attractor solutions in the phase space is developed for the model by simultaneously best fitting the stability and model parameters with the observational data. The results show that for an accelerating universe the phantom crossing does not occur in the past and near future.

\end{abstract}



\maketitle

\section{Introduction}

Recently, the observations of high redshift Type Ia supernovea
(SNe Ia), the surveys of clusters of galaxies \cite{Reiss}--\cite{Riess2}, Sloan digital sky survey ({\bf
SDSS})~\cite{Abazajian} and Chandra X--ray observatory~\cite{Allen} reveal the universe accelerating expansion. Also the
observations of Cosmic Microwave Background (CMB)
anisotropies \cite{Bennett} indicate that the universe is flat and the total energy
density is very close to the critical one \cite{Spergel}. The observations determines basic cosmological parameters
with high precisions and strongly indicates that the universe
presently is dominated by a smoothly distributed and slowly
varying dark energy (DE) component. A dynamical equation of state ( EoS) parameter that is connected directly to the evolution of the energy density in the universe and indirectly to the expansion of the Universe can be regarded as a suitable parameter to explain the acceleration and the origin of DE \cite{Seljak}--\cite{Setare}. Among all the cosmological models, the scalar-tensor theories have been widely used to explain the late time acceleration of the universe and its relation to the dark energy. For a review see \cite{Sahoo}--\cite{farajollahi}

The authors in \cite{hossein} in a separate work have studied the dynamics of the universe in Chameleonic Brans--Dicke (CBD) cosmological model. They perform stability analysis and investigate the attractor solutions of the CBD cosmology by utilizing the 2-dimensional phase space of the theory. In there, for the matter in the universe assumed to be perfect fluid with the EoS parameters, $\gamma=0$ or $ 1/3$, the analysis gives the corresponding conditions for tracking attractor and determines the universe behavior in the past and future. The authors then study the cosmological parameters such as effective EoS parameter for the model in terms of the stability variables. Depending on the stability parameters, the model predicts phantom behavior, despite the absence of phantom energy. It has been shown that by fitting the observational data to the model for the distance modulus, the scenarios with phantom crossing better fits the data.

In an extension, in this work we study the Chameleonic Generalized Brans--Dicke (CGBD) cosmological model by letting the BD parameter to be function of the BD scalar field. A dynamical BD parameter can be favored by realizing that general relativity is recovered from BD model in the limit $\omega\rightarrow \infty$ and large values of the BD scalar field. Also, in a different approach, to perform stability analysis, we simultaneously best fit the model parameters with the observational data using $\chi^2$ method. This enables us to find the best fitted model parameters for the analysis of the critical points and also verification of the model with the experiment.

The paper is organized as follows: Sec. two is devoted to a detailed formulation of the CGBD cosmological model. In Sec. three, we perform stability analysis of the model. We first obtain the autonomous differential equations of the model in terms of dimensionless dynamical variables. We then solve the equations by simultaneously best fitting the model parameters and initial conditions with the data and study the attracting behavior of the critical points in the phase plane. In Sec. four, we examine the behavior of the EoS parameter of the model. In Sec. five, we present a discussion.

\section{The model}

We consider the CGBD gravity in the presence of matter with  the action given by,

\begin{eqnarray}\label{ac}
S=\int d^{4}x\frac{\sqrt{-g}}{2}[\phi
R-\frac{\omega(\phi)}{\phi}\phi^{,\alpha}\phi_{,\alpha}-2V(\phi)+2L_{m}f(\phi)],
\end{eqnarray}
where $R$ is the Ricci scalar, $\phi$ is the Chameleon and/or BD scalar field with the potential $V(\phi)$, and
$\omega(\phi)$ is $BD$ parameter. We assume that $f(\phi)=f_{0}\phi^\kappa$ and $V(\phi)=V_{0}\phi^\delta$ where $\kappa$ and $ \delta $ are  dimensionless constants characterizing the slope of potential $V(\phi)$ and $f(\phi)$. The cosmological models with such exponential functions have been known lead to interesting physics
in a variety of context, ranging from existence of accelerated expansions \cite{Halliwell} to cosmological
scaling solutions \cite{Ratra}--\cite{Yokoyama}. In particular, the exponential forms of $f(\phi)$ and $V(\phi)$ are motivated by chameleon models \cite{Brax2} and also from stability considerations \cite{stability}. In addition, attractor solutions with exponential functions may lead to cosmic acceleration for natural values of model parameters \cite{Barreiro}. Unlike the usual Einstein--Hilbert action, the matter lagrangian
$L_{m}$ is modified as $f(\phi)L_{m}$, where $f(\phi)$ is an
analytic function of the scalar field. The last term in the action
brings about the nonminimal interaction between
matter and the scalar field. Variation of the action (\ref{ac})
with respect to the metric tensor $g_{\mu\nu}$ gives,
\begin{eqnarray}\label{fieldeq}
G_{\mu\nu}=\frac{\omega(\phi)}{\phi^{2}}[\phi_{,\mu}\phi_{,\nu}-\frac{1}{2}g_{\mu\nu}\phi_{,\alpha}\phi^{,\alpha}]
+\frac{1}{\phi}[\phi_{,\mu;\nu}-g_{\mu\nu}\Box \phi]-\frac{V(\phi)}{\phi}g_{\mu\nu}+\frac{f(\phi)}{\phi}T_{\mu\nu}.
\end{eqnarray}
In FRW cosmology, the field equation (\ref{fieldeq}) becomes,
\begin{eqnarray}\label{fried1}
3H^2=\frac{\rho_{m}f(\phi)}{\phi}-3H\frac{\dot{\phi}}{\phi}
+\frac{\omega(\phi)}{2}\frac{\dot{\phi}^{2}}{\phi^{2}}-\frac{V(\phi)}{\phi},
\end{eqnarray}
\begin{eqnarray}\label{fried2}
2\dot{H}+3H^2=-\frac{p_{m}f(\phi)}{\phi}-2H\frac{\dot{\phi}}{\phi}
-\frac{\omega(\phi)}{2}\frac{\dot{\phi}^{2}}{\phi^{2}}-\frac{\ddot{\phi}}{\phi}-\frac{V(\phi)}{\phi}.
\end{eqnarray}
On the other hand, the variation of the action (\ref{ac}) with respect to the spatially homogeneous scalar fields $\phi(t)$ gives,
\begin{eqnarray}\label{phiequation}
\ddot{\phi}+3H\dot{\phi}=\frac{(\rho_{m}-3p_{m})f(\phi)}{3+2\omega(\phi)}
-\frac{2(2V(\phi)-\phi V')}{3+2\omega(\phi)}
-\frac{(\rho_{m}-3p_{m})\phi f'}{2(3+2\omega(\phi))}-\frac{\omega^{'}\dot{\phi}^{2}}{3+2\omega(\phi)},
\end{eqnarray}
where prime " $^\prime$ " means derivative with respect to $\phi$. We assume that the universe is filled with the barotropic fluid
with the EoS, $p_{m}=\gamma\rho_{m}$. From equations (\ref{fried1}), (\ref{fried2}) and (\ref{phiequation}), one can easily arrive at the modified conservation equation,
\begin{eqnarray}
\dot{(\rho_{m}f)}+3\frac{\dot{a}}{a}(1+\gamma)\rho_{m}f=\frac{1}{4}(1-3\gamma)\rho_{m}\dot{f}.
\end{eqnarray}
In the next section we study the stability analysis of the model in the phase space.

\section{perturbation and Stability Analysis}

In this section, we study the structure of the dynamical system via  phase plane analysis,
by introducing the following dimensionless variables,
\begin{eqnarray}\label{defin}
 \chi^{2}={\frac{\rho_{m}f}{3\phi H^{2}}},\ \ \zeta={\frac{\dot{\phi}}{\phi H}} ,\ \ \eta^{2}=\frac{V}{3\phi H^{2}}.
\end{eqnarray}
Using equations (\ref{fried1})-(\ref{phiequation}), the evolution equations of these variables become,
\begin{eqnarray}\label{kai}
\chi^\prime&=&\chi\{\frac{-3(1+\gamma)}{2}+[\frac{\kappa(1-3\gamma)-20}{8}-\frac{\beta \omega_{0}e^{\alpha N}}{6+4\omega_{0}e^{\alpha N}}]\zeta+\frac{\omega_{0}e^{\alpha N}}{2}\zeta^{2}
\\&+&[\frac{3(1-3\gamma)(2-\kappa)}{4(3+2\omega_{0}e^{\alpha N})}+
\frac{3(1+\gamma)}{2}]\chi^{2}+\frac{3\delta-6}{3+2\omega_{0}e^{\alpha N}}\eta^{2})\},\ \nonumber\\
\zeta^\prime&=&\zeta\{-3-3\zeta-\frac{\alpha \omega_{0}e^{\alpha N}}{6+4\omega}\zeta+\frac{\omega_{0}e^{\alpha N}}{2}\zeta^{2}+[\frac{3(1-3\gamma)(2-\kappa)}{4(3+2\omega_{0}e^{\alpha N})}+
\frac{3(1+\gamma)}{2}]\chi^{2}\\&+&\frac{3\delta-6}{3+2\omega_{0}e^{\alpha N}}\eta^{2}\}+\frac{3(1-3\gamma)(2-\kappa)}{2(3+2\omega_{0}e^{\alpha N})}\chi^{2}+\frac{6\delta-12}{3+2\omega_{0}e^{\alpha N}}\eta^{2},\ \nonumber\\
\eta^\prime&=&\eta\{\frac{\delta-5}{2}\zeta-\frac{\alpha \omega_{0}e^{\alpha N}}{6+4\omega_{0}e^{\alpha N}}\zeta+\frac{\omega_{0}e^{\alpha N}}{2}\zeta^{2}+[\frac{3(1-3\gamma)(2-\kappa)}{2(3+2\omega_{0}e^{\alpha N})}+ \label{zeta}
\frac{3(1+\gamma)}{2}]\chi^{2}\\&+& \nonumber\frac{3\delta-6}{3+2\omega_{0}e^{\alpha N}}\eta^{2}\},
\end{eqnarray}
where prime " $^\prime$ " here and from now on is taken to be derivative with respect to $N = ln (a)$. The parameter $\alpha$ in the above equation is defined by  $\alpha\equiv\frac{\dot{\omega}}{\omega H}$ which gives $ \omega\equiv\omega_{0}e^{\alpha N}$. Also, the Friedmann constraint equation (\ref{fried1}) in terms of the new dynamical variables becomes
\begin{eqnarray}\label{constraint}
\chi^{2}-\zeta+\frac{\omega_{0}e^{\alpha N}}{6}\zeta^{2}-\eta^{2}=1.
\end{eqnarray}
In term of the these variables we also obtain,
\begin{eqnarray}
\frac{\dot{H}}{H^{2}}=(2+\frac{\alpha\omega_{0}e^{\alpha N}}{6+4\omega_{0}e^{\alpha N}})\zeta-\frac{\omega_{0}e^{\alpha N}}{2}\zeta^{2}-[\frac{3(1-3\gamma)(2-\kappa)}{4(3+2\omega_{0}e^{\alpha N})}-
\frac{3(1+\gamma)}{2}]\chi^{2}-\frac{3\delta-6}{3+2\omega_{0}e^{\alpha N}}\eta^{2}.\label{hhdot}
\end{eqnarray}
Using the Friedmann constraint (\ref{constraint}), the three equations (\ref{kai})-(\ref{zeta}) reduce to the following two equations:
\begin{eqnarray}
\chi^\prime&=&\chi\{\frac{-3(1+\gamma)}{2}+[\frac{\kappa(1-3\gamma)-20}{8}-\frac{\alpha \omega_{0}e^{\alpha N}}{6+4\omega_{0}e^{\alpha N}}]\zeta+\frac{\omega_{0}e^{\alpha N}}{2}\zeta^{2}
\label{kai1}\\&+&[\frac{3(1-3\gamma)(2-\kappa)}{4(3+2\omega_{0}e^{\alpha N})}+
\frac{3(1+\gamma)}{2}]\chi^{2}+\frac{3\delta-6}{3+2\omega_{0}e^{\alpha N}}(-1+\chi^{2}-\zeta+\frac{\omega_{0}e^{\alpha N}}{6}\zeta^{2})\},\ \nonumber\\
\zeta^\prime&=&\zeta\{-3-3\zeta-\frac{\alpha \omega_{0}e^{\alpha N}}{6+4\omega_{0}e^{\alpha N}}\zeta+\frac{\omega_{0}e^{\alpha N}}{2}\zeta^{2}+[\frac{3(1-3\gamma)(2-\kappa)}{4(3+2\omega_{0}e^{\alpha N})}+\label{zeta1}
\frac{3(1+\gamma)}{2}]\chi^{2}\\&+&\ \nonumber \frac{3\delta-6}{3+2\omega_{0}e^{\alpha N}}(-1+\chi^{2}-\zeta+\frac{\omega_{0}e^{\alpha N}}{6}\zeta^{2})\}
+\frac{3(1-3\gamma)(2-\kappa)}{4(3+2\omega_{0}e^{\alpha N})}\chi^{2}\\&+& \nonumber\frac{6\delta-12}{3+2\omega_{0}e^{\alpha N}}(-1+\chi^{2}-\zeta+\frac{\omega_{0}e^{\alpha N}}{6}\zeta^{2}).\ \nonumber
\end{eqnarray}
It is more convenient to investigate the properties of the dynamical system equations(\ref{kai1}) and (\ref{zeta1})
than equations (\ref{kai})-(\ref{zeta}). We, now, obtain the critical points  and study
their stability of these points. Critical points are always exact constant solutions in the
context of autonomous dynamical systems. These points are often the extreme points of
the orbits and therefore describe the asymptotic behavior of the system. In the following, we find fixed points by simultaneously solving $\chi^\prime=0$ and $\zeta^\prime$. Substituting
linear perturbations $\chi^\prime\rightarrow \chi^\prime+\delta \chi^\prime$, $\zeta^\prime\rightarrow \zeta^\prime+\delta \zeta^\prime$ about the critical points into the two independent equations (\ref{kai1}) and (\ref{zeta1}), to the first
orders in the perturbations, one yields two eigenvalues $\lambda_{i} (i=1,2)$. Stability requires the real part of the eigenvalues to be negative.

In the following, the above equations can be solved to give fixed points (critical points) for two scenarios of the matter field, $\gamma=0$ and  $\gamma=1/3$. Both critical points and eigenvalues in our model are highly nonlinear and depend on the stability parameters $\kappa$, $\delta$ and model parameters $\alpha$ and $\omega_0$. In addition, the expressions for the critical points and eigenvalues are long and cumbersome such that it is not easy to find out under what conditions the critical points are stable or unstable.

In a different approach we solve the above equations by best fitting the stability and model parameters and initial conditions with the observational data for distance modulus using the $\chi^2$ method. This helps us to find the solutions for the above equations and conditions for the stability of the critical points that are physically more meaningful and observationally more favored. Next, we examine this idea of simultaneously fitting the model and solving the field equations.

\subsection{Best fitting the stability parameters and initial conditions}

The difference between the absolute and
apparent luminosity of a distance object is given by, $\mu(z) = 25 + 5\log_{10}d_L(z)$ where $d_L(z)$ is the Luminosity distance quantity. If we  consider a possible variation of the
effective gravitational constant, $G_{eff}$, in the Universe acceleration
rate history, an additional term made of
with the ratio between the value of effective gravitational
constant at any redshift and the same quantity evaluated
at the present has to be contributed, \cite{Riazuelo}
\begin{equation}\label{dl}
\mu(z;\beta,L;\lambda) = 25 + 5\log_{10}d_L(z)+\frac{15}{4}\log\frac{G_{eff}(z;\beta,L;\lambda)}{G_{eff}(0;\beta,m;k)},
 \end{equation}
where
\begin{equation}\label{dl}
G_{eff}(z;\beta,L;\lambda) = G_N(1+2\beta\frac{(1+z)62L^2}{(1+z)^2L^2+\lambda^2}),\nonumber
 \end{equation}
with $G_N$ is the Newton's gravitational constant. The intrinsic model parameters, i.e., the coupling constant $\beta$ and the interaction length $L\propto m^{-1}$ can be constrained with the fitting procedure. The length $\lambda\propto k^{-1}$ (wavelength $k$) for the SNe Ia redshift range is of order $10^{-2}$. As explained in \cite{Riazuelo}, a time-varying gravitational constant can affect light curves from SNe Ia. However, the correction coming from it is negligible when compared to the usual distance modulus expression.

From numerical computation one can obtain $H(z)$ which can be used to evaluate $\mu(z)$. To best fit the model for the stability and model parameters $\delta$, $\kappa$, $\alpha$ and $\omega_{0}$ and the initial conditions $\chi(0)$, $\zeta(0)$, $H(0)$ with the most recent observational data, the SNe Ia, we employe the $\chi^2$ method. In addition, from \cite{mota} we use the best fitted values obtained for the parameters $\beta$ and $L$ in calculating the last term of the equation (\ref{dl}). We constrain the parameters and also the initial conditions by minimizing the $\chi^2$ function given by
\begin{equation}\label{chi2}
    \chi^2_{SNe}(\delta, \kappa,\alpha,\omega_{0}, \chi(0), \zeta(0), H(0))=\sum_{i=1}^{557}\frac{[\mu_i^{the}(z_i|\delta, \kappa,\alpha,\omega_{0}, \chi(0), \zeta(0), H(0)) - \mu_i^{obs}]^2}{\sigma_i^2},
\end{equation}
where the sum is over the SNe Ia sample. In the relation (\ref{chi2}), $\mu_i^{the}$ and $\mu_i^{obs}$ are the distance modulus parameters obtained from our model and from observation, respectively, and $\sigma$ is the estimated error of the $\mu_i^{obs}$. From numerical computation in table I, we show the best fitted model parameters for $\gamma=0$ and $\gamma=1/3$.\\

\begin{table}[ht]
\caption{Best fitted values for the Stability parameters, model parameters and Initial conditions.} 
\centering 
\begin{tabular}{c c c c c c c c c} 
\hline\hline 
parameters  &  $\delta$  & $\kappa$ \ & $\alpha$ \ & $\omega_{0}$\ & $\chi(0)$\ & $\zeta(0)$\ & $H(0)$ \ & $\chi^2_{min}$\\ [4ex] 
\hline 
$\gamma=0$&$-3.1$  & $22$ \ & $-1.3$ \ & $9.7$\ & $-0.3$\ & $0.2$\ & $0.91$ \ & $543.1628893$ \\
$\gamma=1/3$  & $-0.04$  & $-$ \ & $-1.78$ \ & $6.2$\ & $-0.3$\ & $0.1$\ & $0.91$ \ & $552.4495936$ \\
\hline 
\end{tabular}
\label{table:1} 
\end{table}\

Fig. 1) shows the constraints on the parameters $\delta$, $\kappa$, $\alpha$ and $\omega_{0}$ and the initial conditions $\chi(0)$, $\zeta(0)$ and $H(0)$ at the $68.3\%$, $95.4\%$ and $99.7\%$ confidence levels in both cases of $\gamma=0, 1/3$. Note that in case of $\gamma=1/3$, since in equation (\ref{hhdot}) the third term vanishes, no constraint on $\kappa$ is expected, as can be seen in Fig.1.

\begin{tabular*}{2.5 cm}{cc}
\includegraphics[scale=.3]{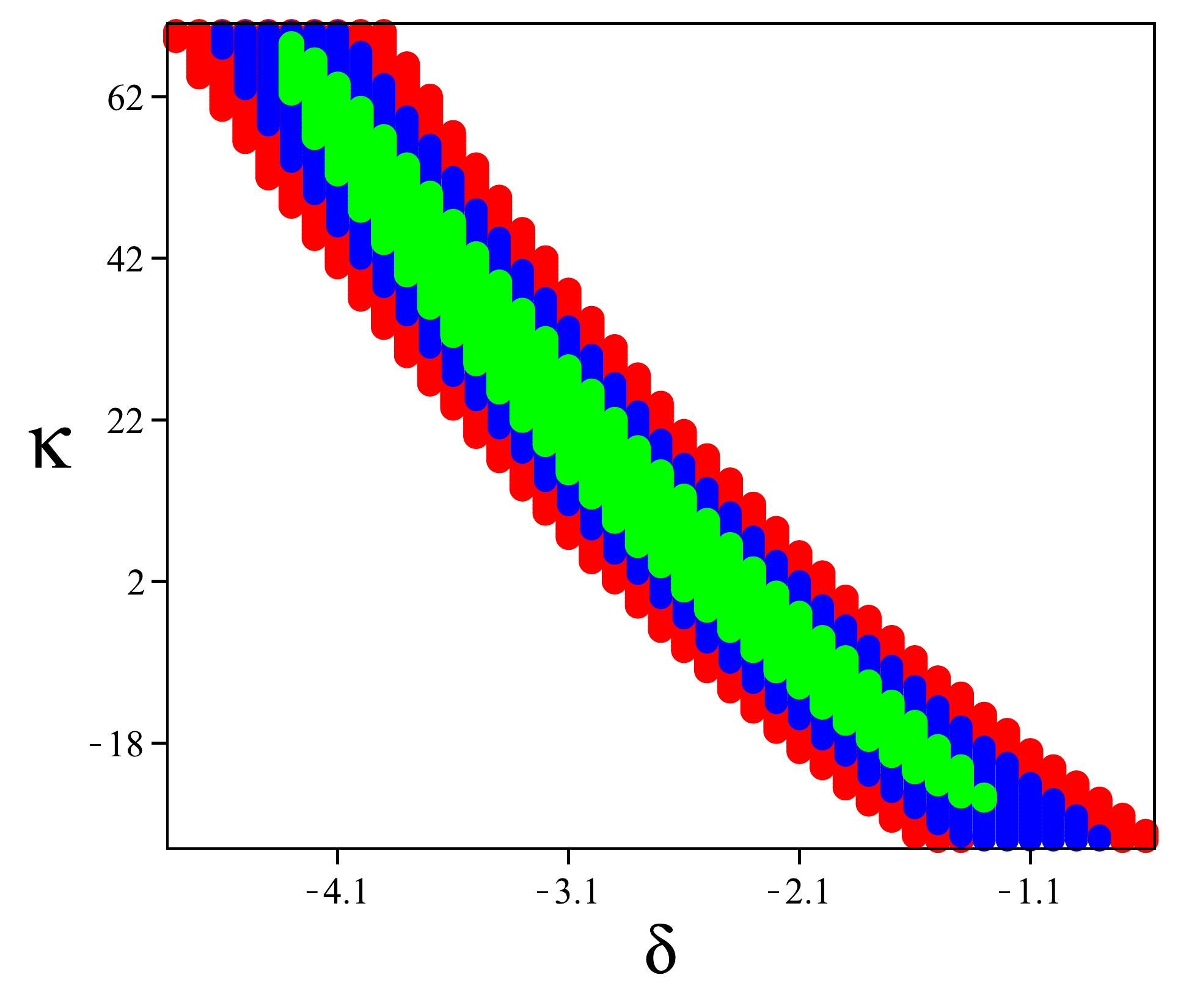}\hspace{0.1 cm}\includegraphics[scale=.3]{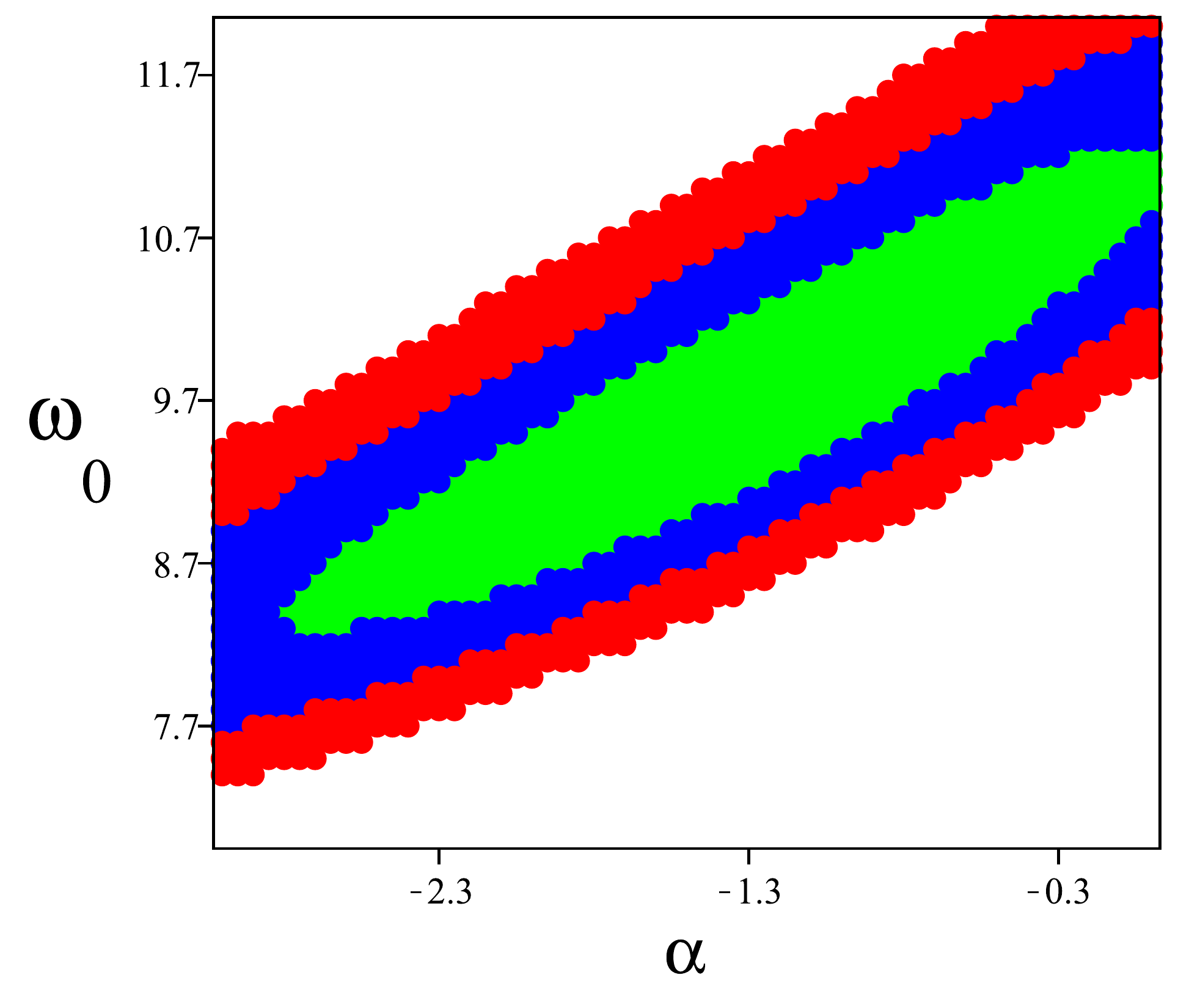}\hspace{0.1 cm}\includegraphics[scale=.26]{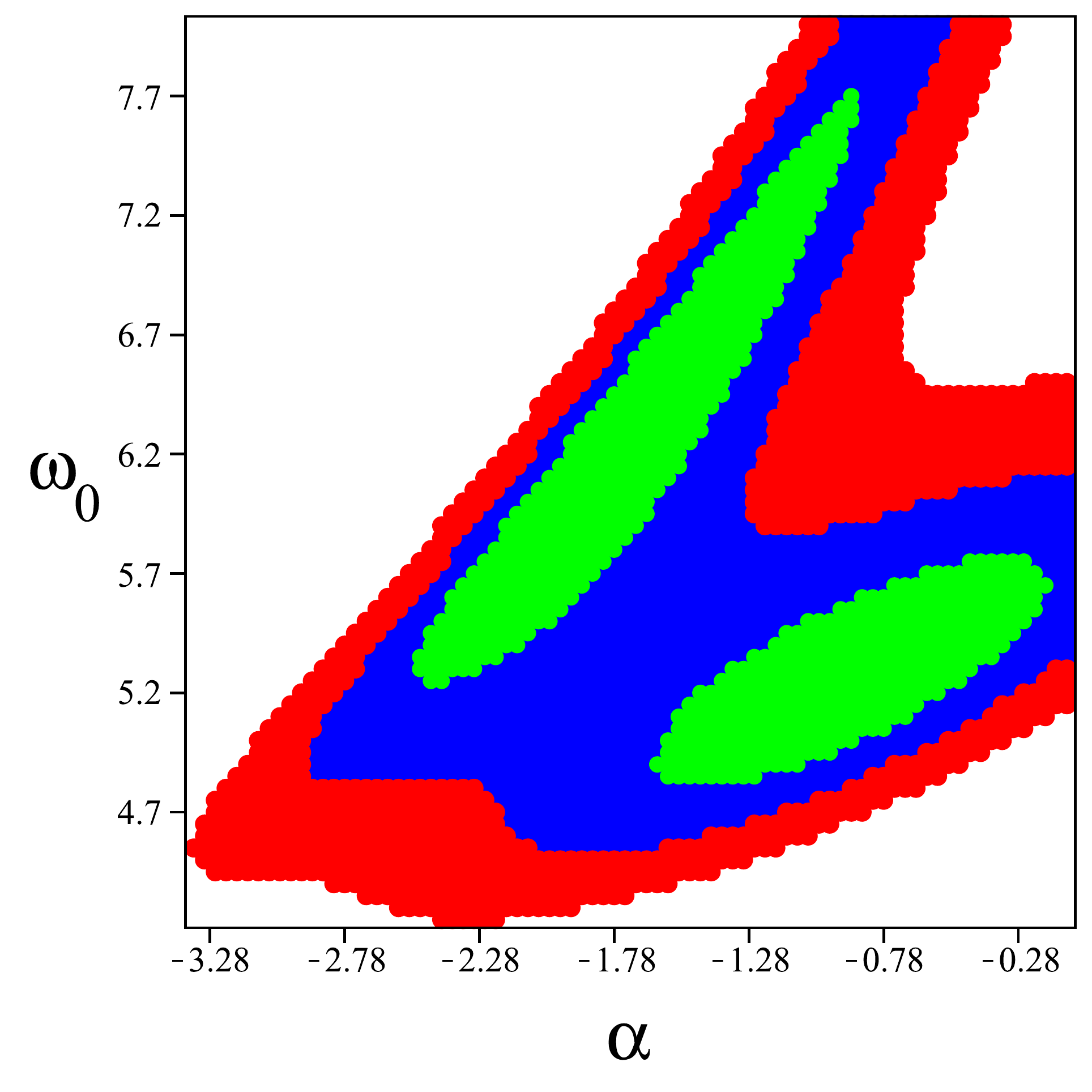}\hspace{0.1 cm}\\
Fig. 1:  The constraints at the 68.3\%, 95.4\% and 99.7\% confidence levels \\from Sne Ia for the parameters  $\delta$, $\kappa$, $\alpha$ and $\omega_{0}$\\
 and the I.C.s $\chi(0)$, $\zeta(0)$ and $H(0)$ in case of: left $\&$ middle) $\gamma=0$
right) $\gamma=1/3$
\end{tabular*}\\
Alternatively, we can plot the likelihood for the model parameters in both cases $\gamma=0$ and $\gamma=1/3$ ( Figs. 2 and 3).

\begin{tabular*}{2.5 cm}{cc}
\includegraphics[scale=.3]{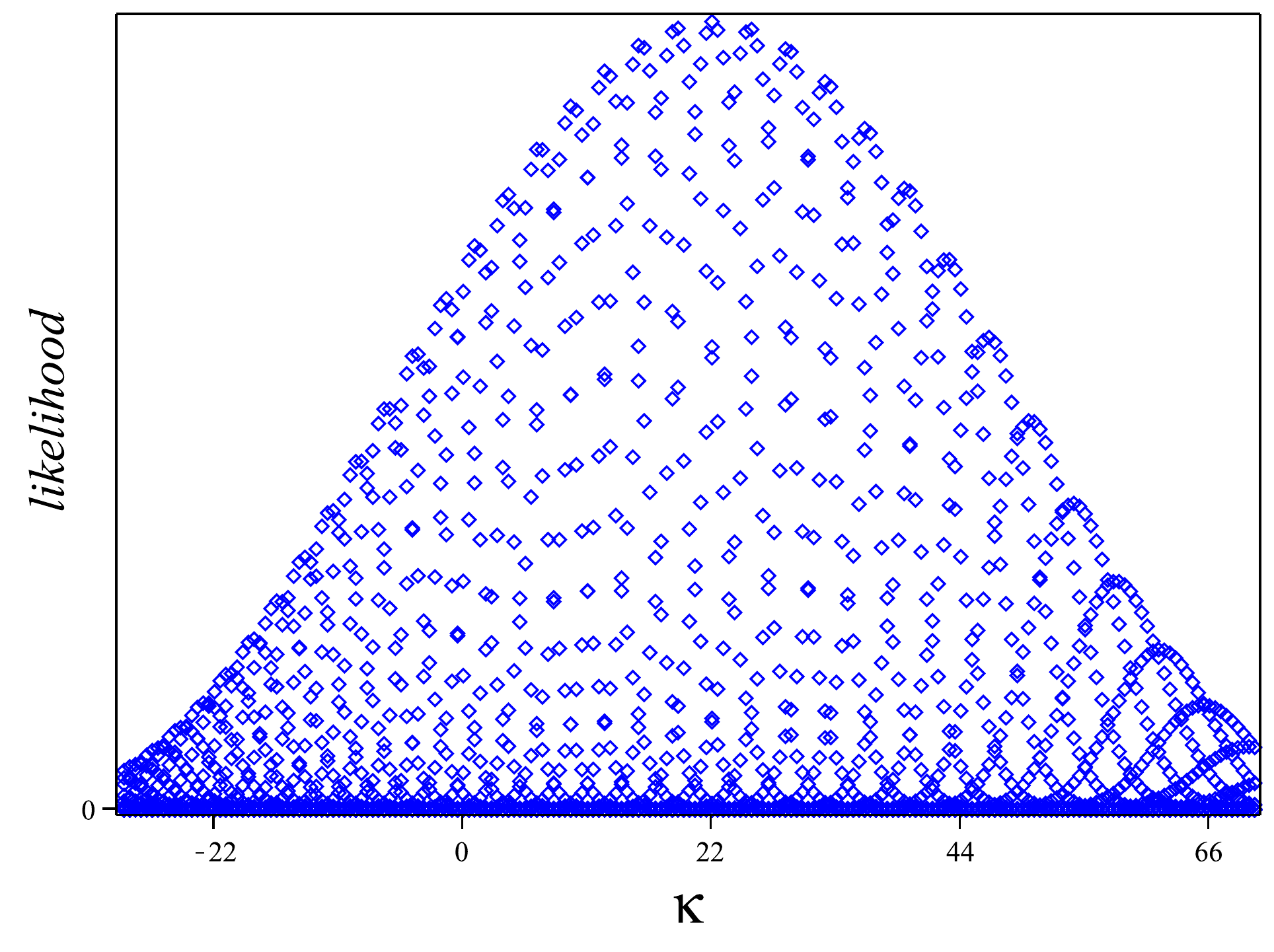}\hspace{0.1 cm}\includegraphics[scale=.3]{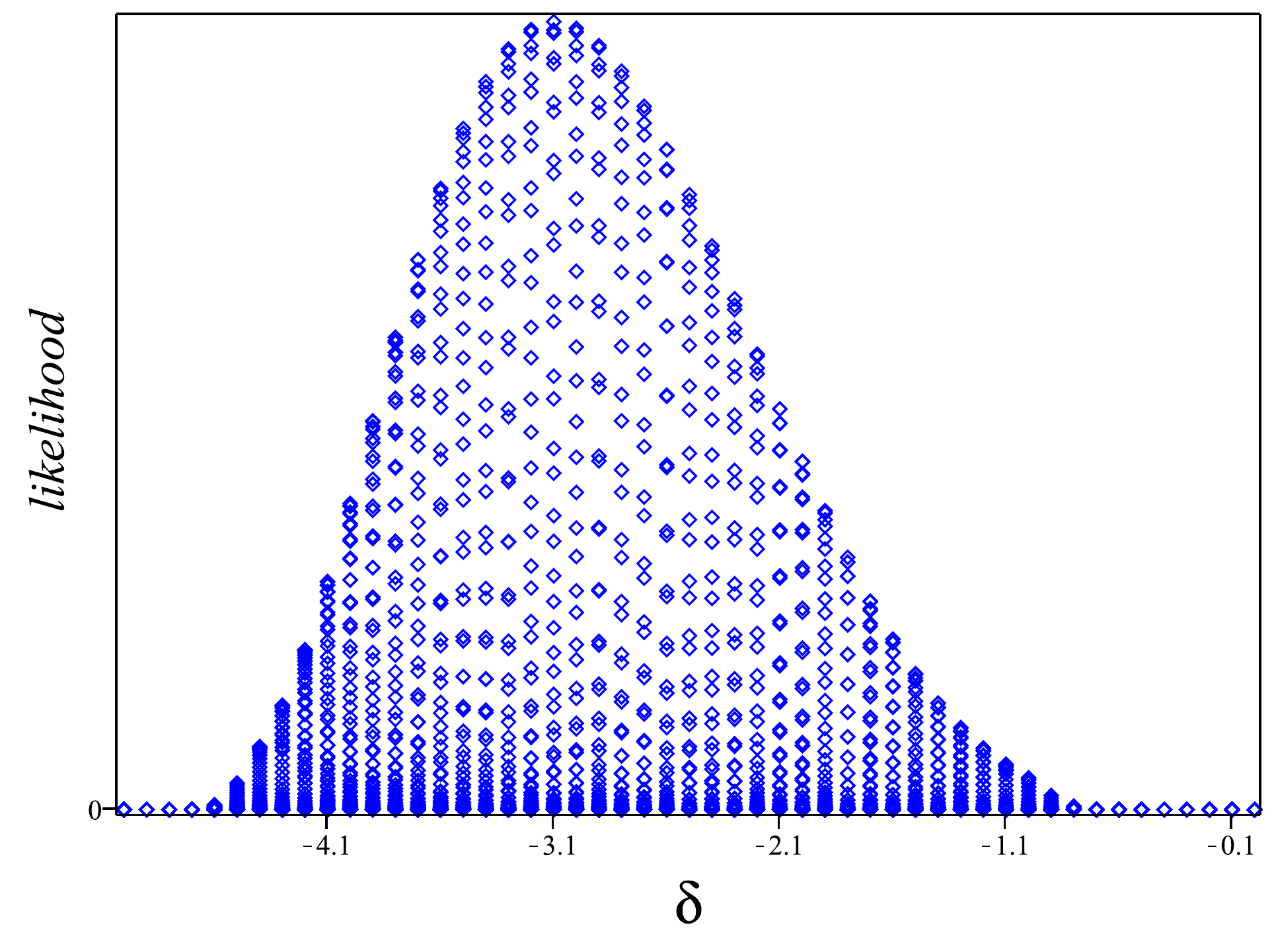}\hspace{0.1 cm}\includegraphics[scale=.3]{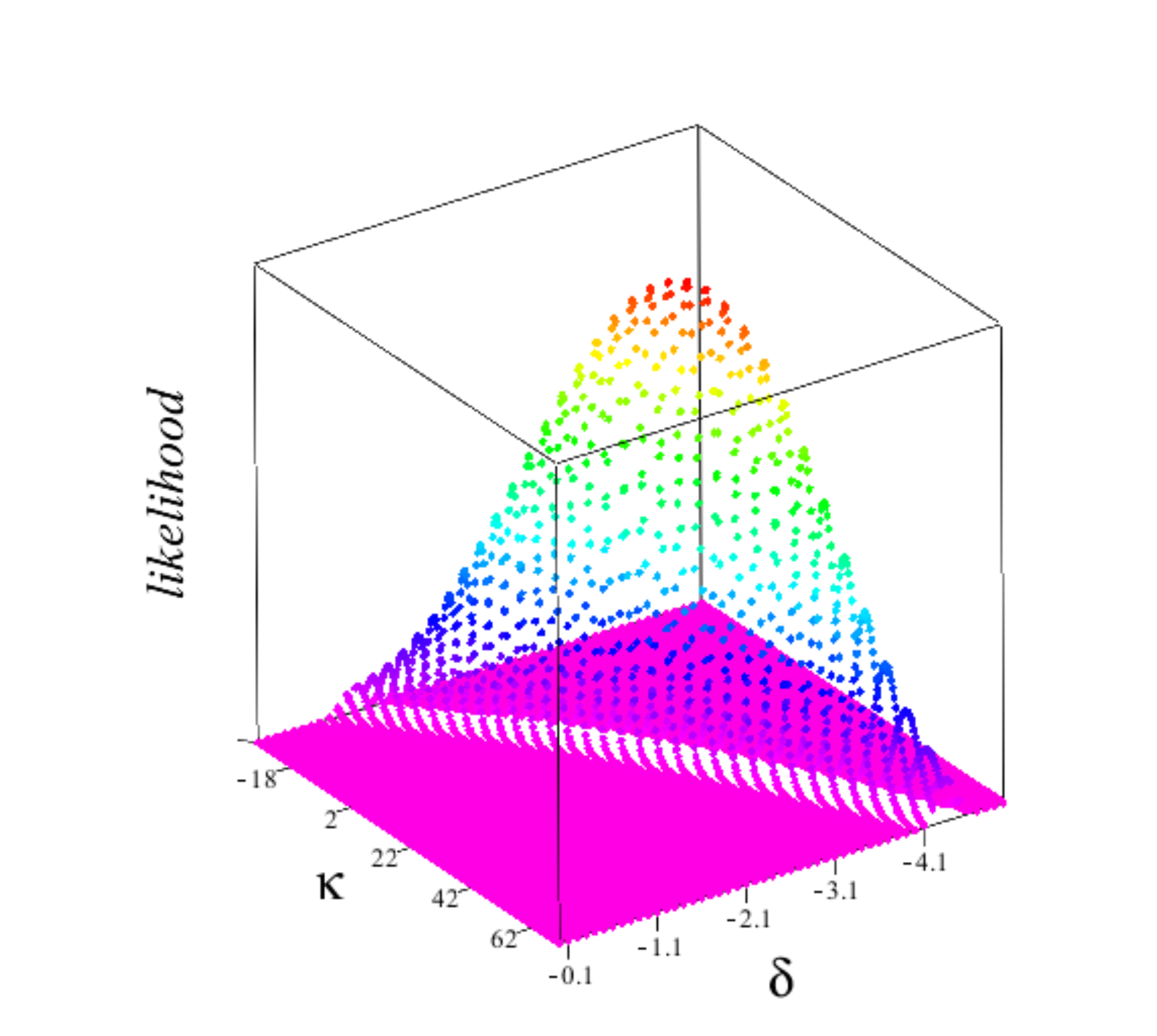}\hspace{0.1 cm}\\
\includegraphics[scale=.3]{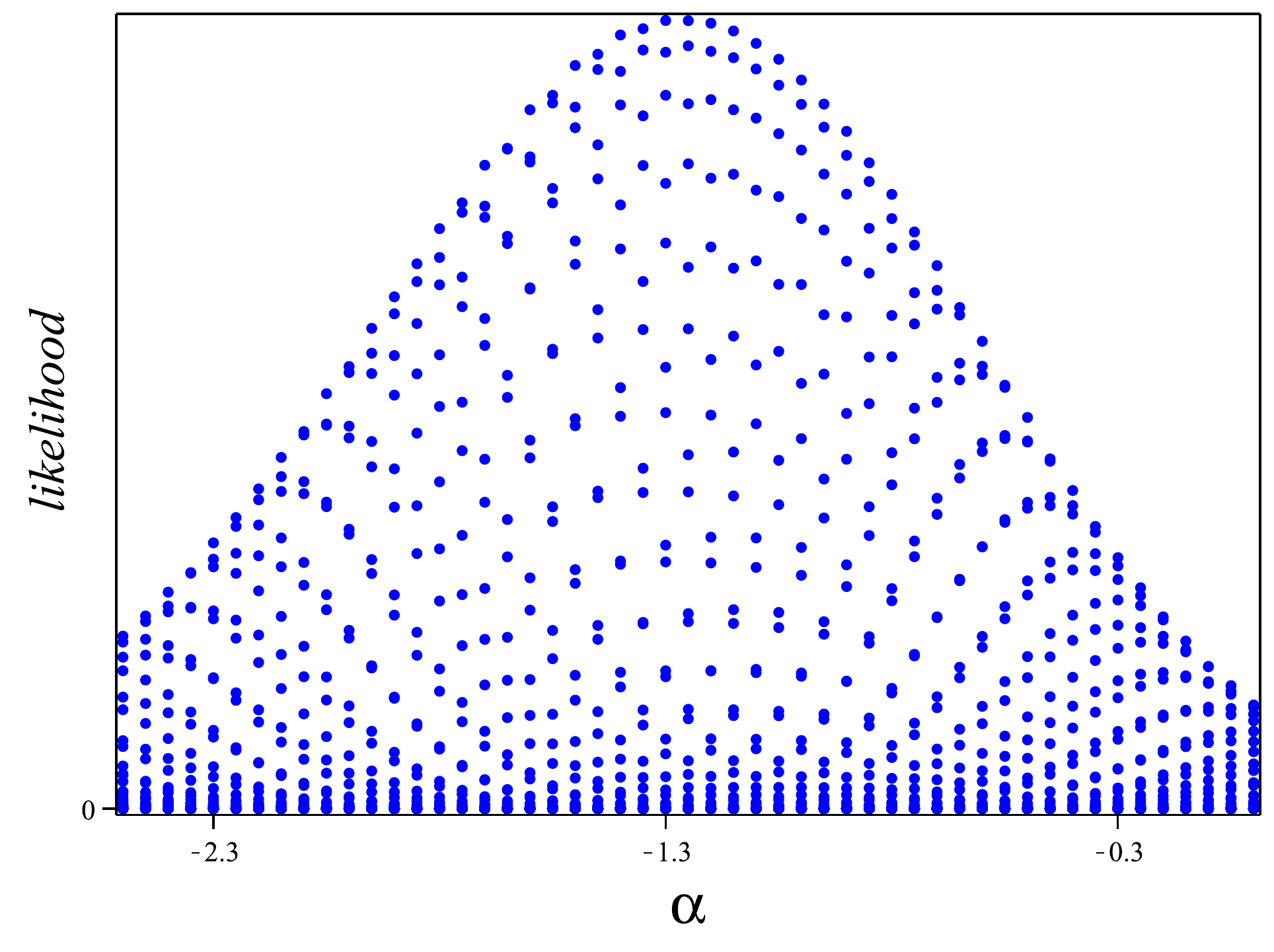}\hspace{0.1 cm}\includegraphics[scale=.3]{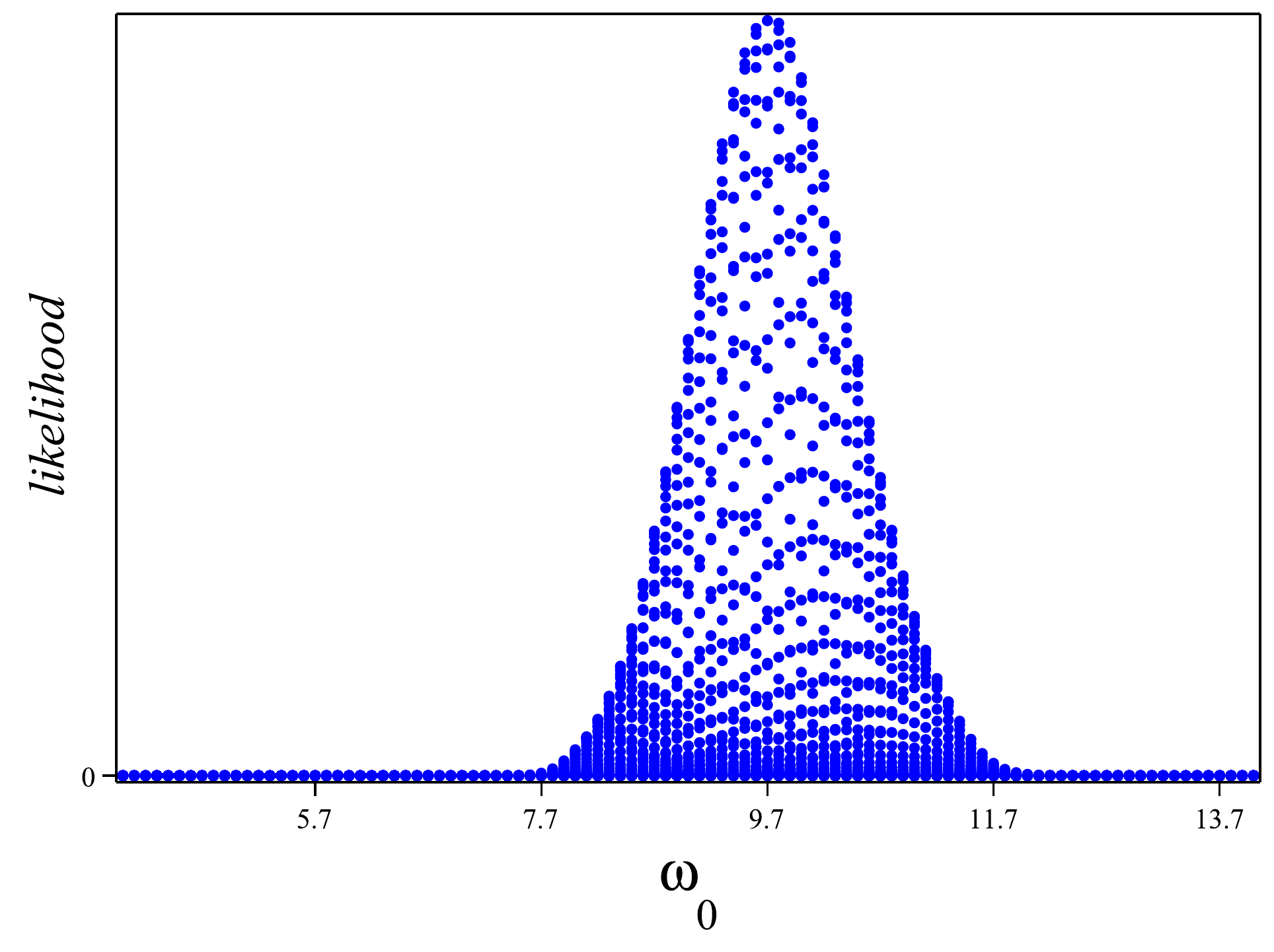}\hspace{0.1 cm}\includegraphics[scale=.3]{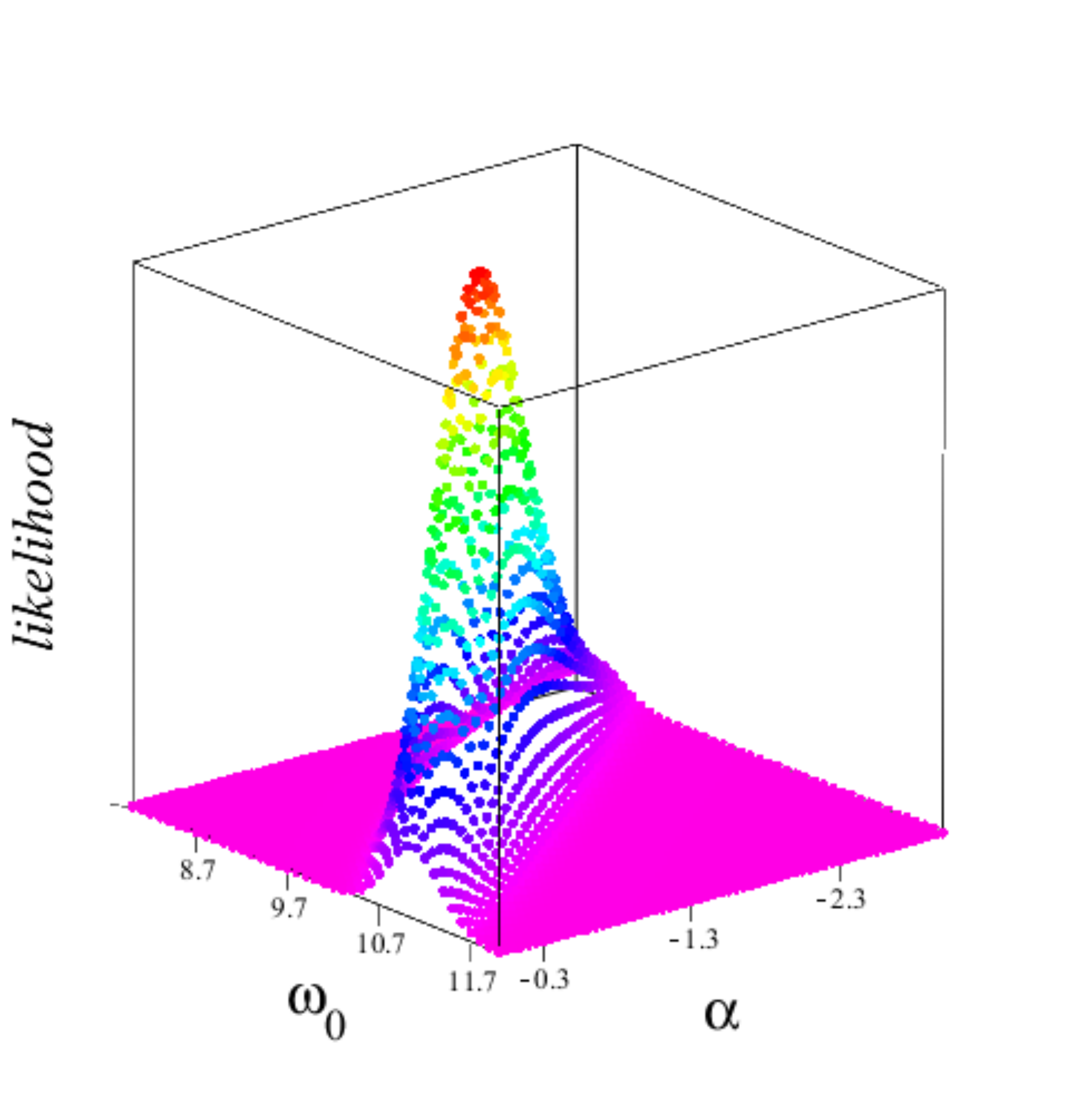}\hspace{0.1 cm}\\
Fig. 2: The 1-dim and 2-dim likelihood distribution for parameters $\kappa$, $\delta$, $\omega_{0}$ and $\alpha$ in $\gamma=0$ case.\\
\end{tabular*}\\

\begin{tabular*}{2.5 cm}{cc}
\includegraphics[scale=.3]{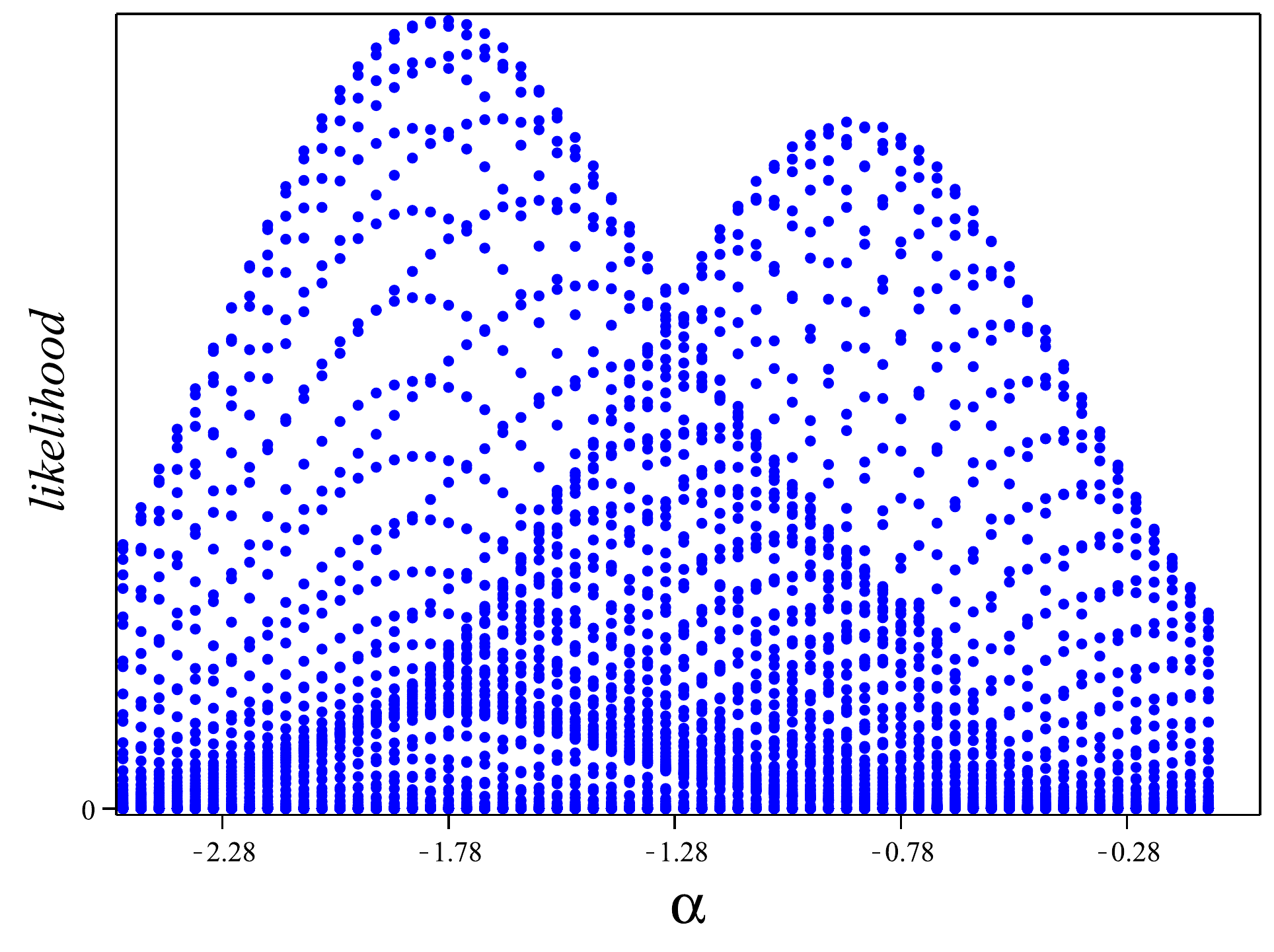}\hspace{0.1 cm}\includegraphics[scale=.3]{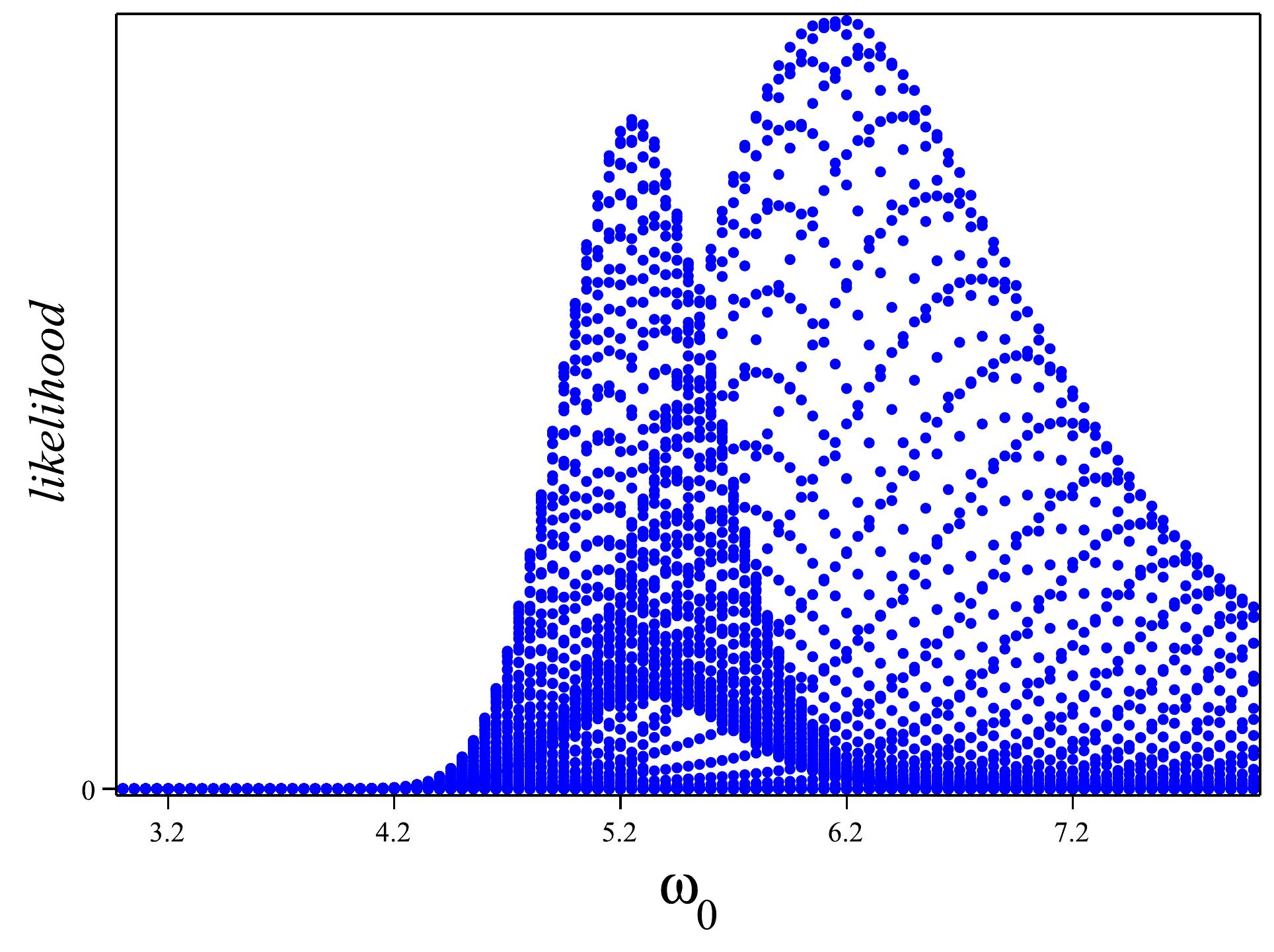}\hspace{0.1 cm}\includegraphics[scale=.3]{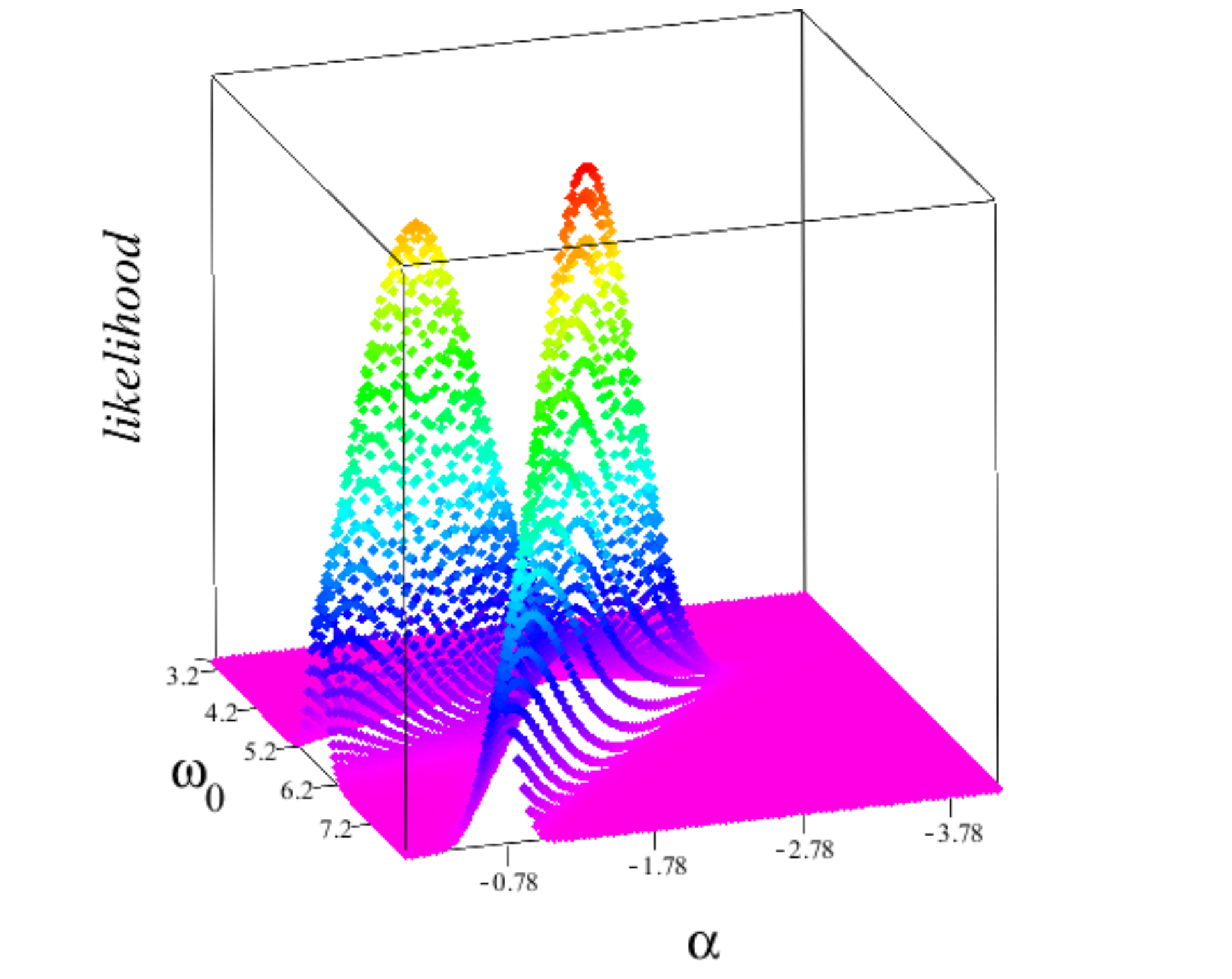}\hspace{0.1 cm}\\
Fig. 3:  The 1-dim and 2-dim likelihood distribution for parameters $\omega_{0}$ and $\alpha$ in $\gamma=1/3$ case. \\
\end{tabular*}\\

The distance modulus, $\mu(z)$, plotted in Fig. 4, in both cases $\gamma = 0, 1/3$ are fitted with the SNe Ia observational data for the model parameters and initial conditions using $\chi^2$ method.

\begin{tabular*}{2.5 cm}{cc}
\includegraphics[scale=.45]{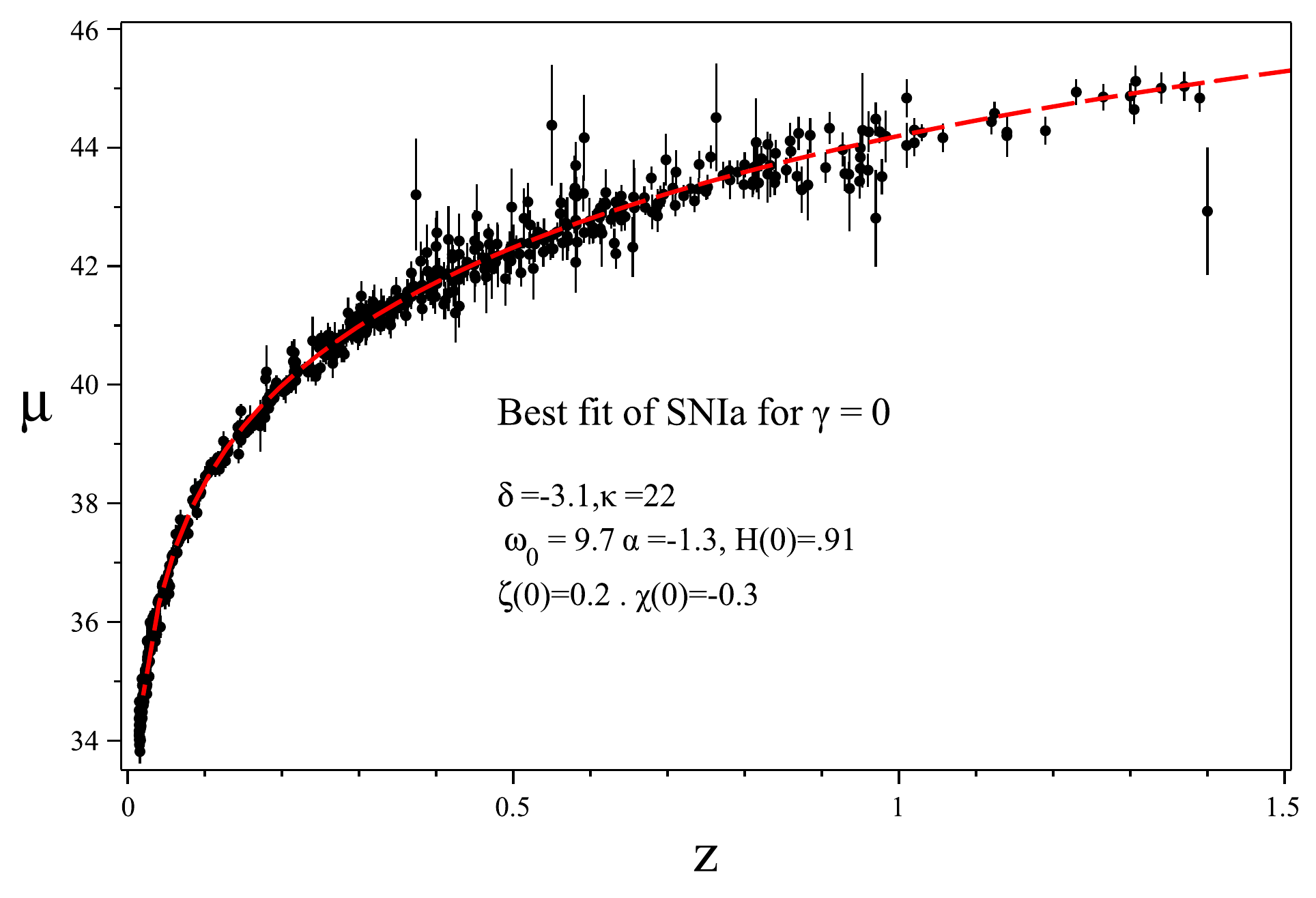}\hspace{0.1 cm}\includegraphics[scale=.45]{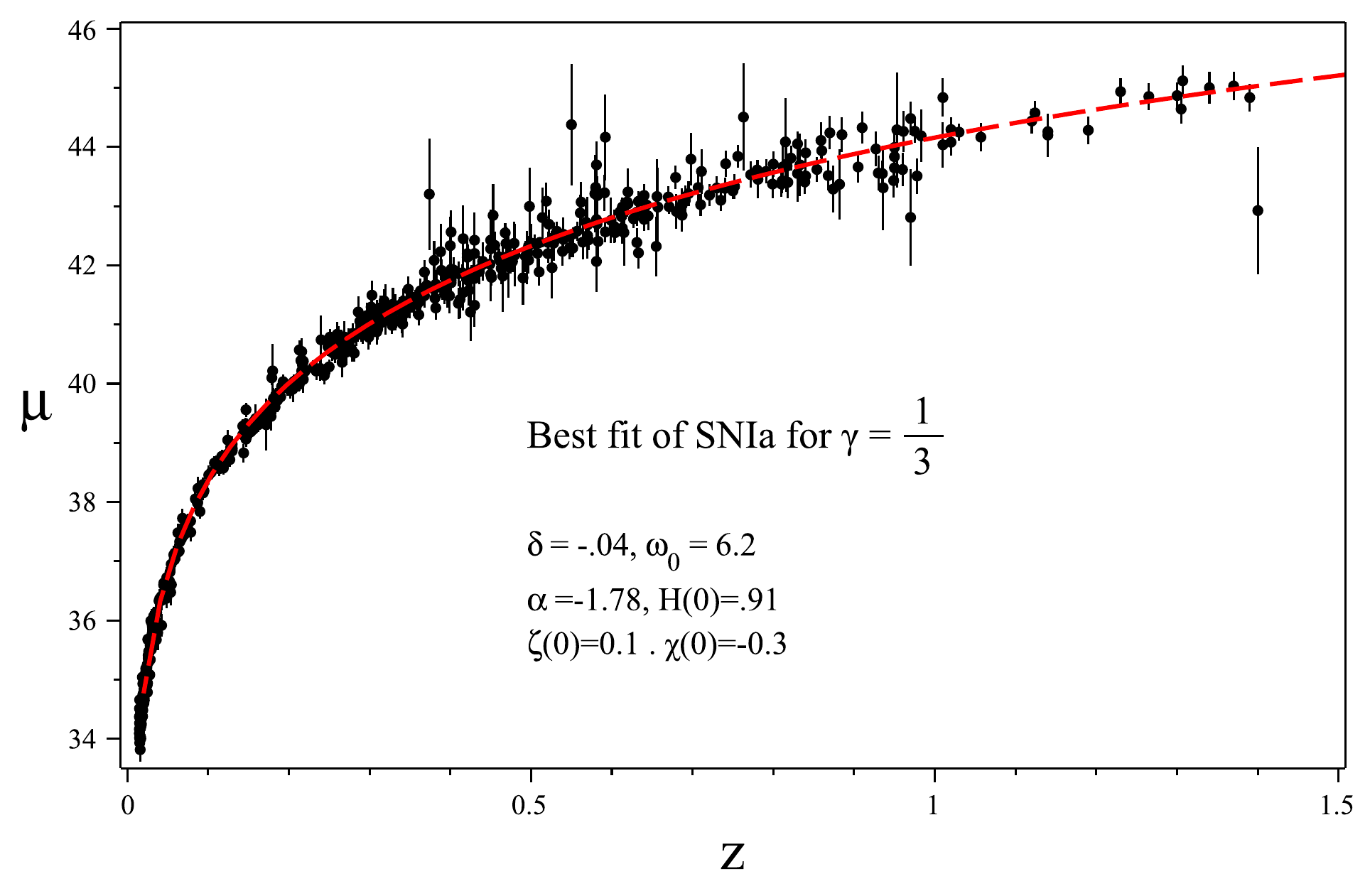}\hspace{0.1 cm}\\
Fig. 4: The distance modulus $\mu(z)$ plotted as function of redshift for the best fitted parameters\\ in both cases $\gamma=0, 1/3$.\\
\end{tabular*}\\

In the following we investigate the stability of the model
with respect to the best fitted model parameters and for the two specific choices of the EoS parameter for the matter in the universe, i.e. $\gamma=0$ and  $\gamma=1/3$.

\subsection{Stability of the best fitted critical points and phase space}

Solving the stability equations for the best fitted model parameters we find fixed points with their stability properties as illustrated in tables II and III, for $\gamma=0$ and $\gamma=1/3$ respectively.\\

\begin{table}[ht]
\caption{Best fitted critical points for $\gamma=0$} 
\centering 
\begin{tabular}{c c c c c } 
\hline\hline 
points  & $(\chi, \zeta)  $  & Stability \\  
\hline 
 P1&(0, 0) & unstable
 \\
 \hline
 P2 & $(0.78, 0.37)$ & stable
 \\
 \hline
 P3 & $(-0.78, .37)$ & stable
 \\
\end{tabular}
\label{table:1} 
\end{table}\

\begin{table}[ht]
\caption{Best fitted critical points for $\gamma=\frac{1}{3}$} 
\centering 
\begin{tabular}{c c c c c } 
\hline\hline 
points  & $(\chi, \zeta)  $  & Stability \\
\hline 
 P1&(0, 0) & unstable
 \\
 \hline
 P2 & (1, 0) & saddle point
 \\
\hline 
 P3&$(-1, 0) $ & saddle point
 \\
 \hline
 P4 & $(0, 4.3)$& stable
 \\
\end{tabular}
\label{table:1} 
\end{table}\

From the above tables we see that, by best fitting the stability parameters, the critical
points P2 and P3 for $\gamma=0$ and P4 for $\gamma=1/3$ are stable and the others are unstable.

In Fig. 5)left), for $\gamma=0$, the trajectories leaving the unstable critical point P1 in the past in the phase plane is shown going towards the stable critical point P2 or p3 in the future. The best fitted model parameter trajectories is also shown by red color dashed trajectory).

 In Fig. 5)right), for $\gamma=1/3$, there are three unstable critical points including two saddle points, P1 and P2.  With the given initial conditions for $\chi(0)$ between -1 and 1, the trajectories (blue and red curves) leaving the unstable critical point P1 in the past in the phase plane is shown moving towards the stable critical point P4 in the future. For the initial conditions less than -1 or greater than 1, then the trajectory shown by green curve leaving the unstable critical point P1 moving towards the saddle points P2 and P3 and leaving these points towards infinity.

\begin{tabular*}{2.5 cm}{cc}
\includegraphics[scale=.45]{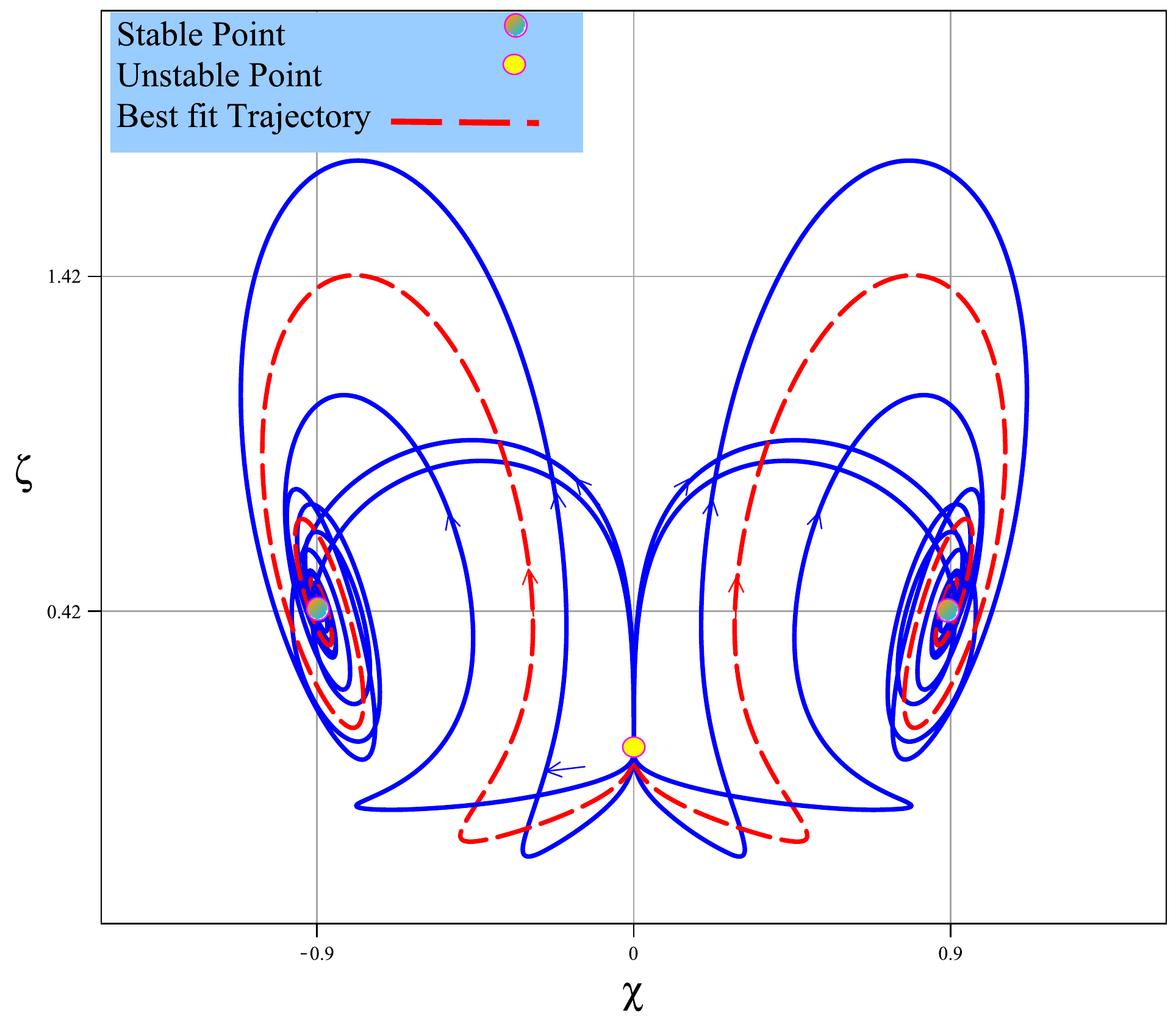}\hspace{0.1 cm}\includegraphics[scale=.45]{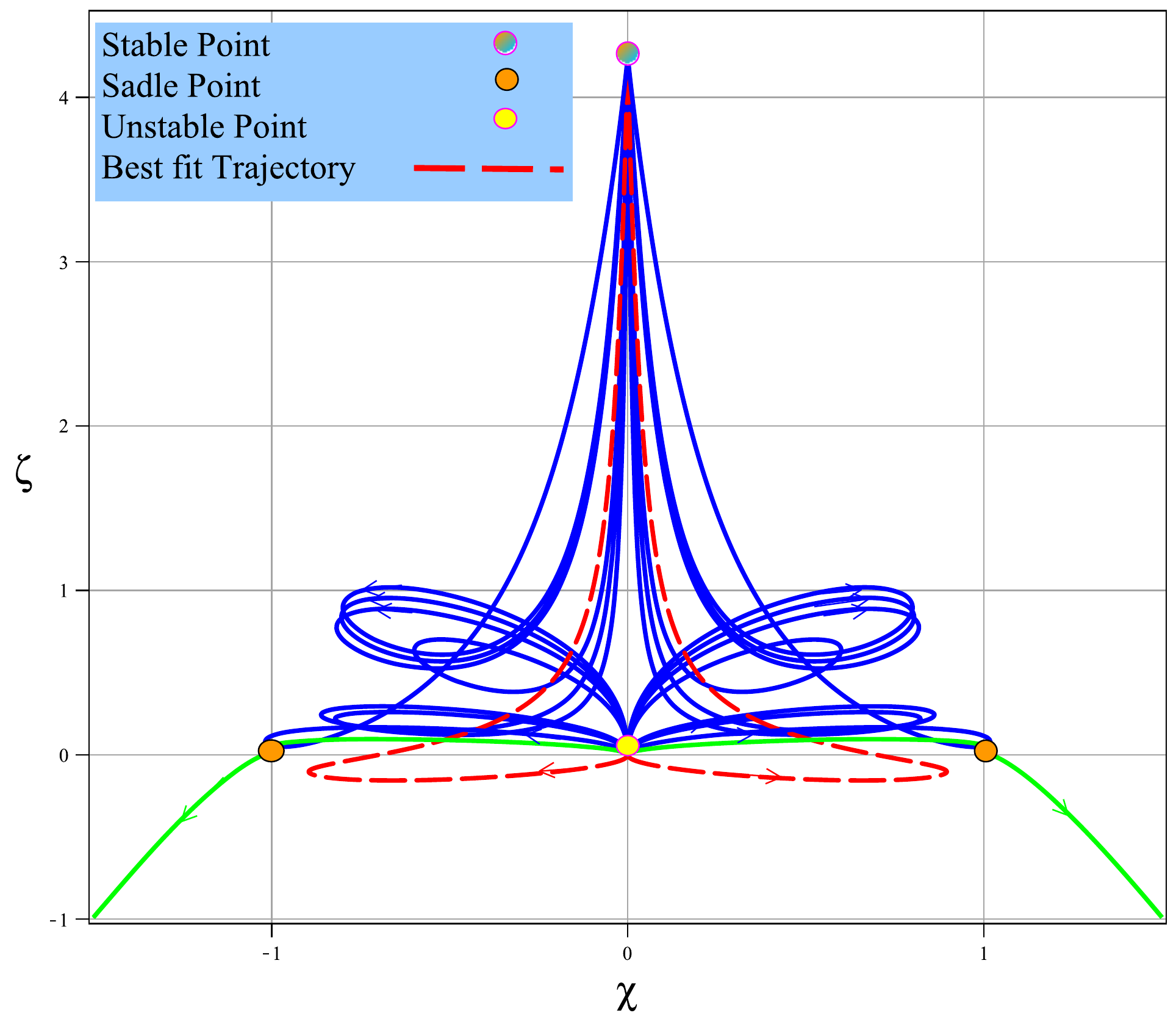}\hspace{0.1 cm}\\
Fig. 5:  The attractor property of the dynamical system in the phase plane for\\ left) $\gamma=0$ and right) $\gamma=1/3$.\\
\end{tabular*}\\

\section{Cosmological parameters: EoS parameter}

In order to understand the behavior of the universe and its dynamics we need to study the cosmological parameters such as EoS parameter. We have already fitted our model with the current observational data for the distance modulus. The EoS parameters analytically and/or numerically have been investigated by many authors for variety of cosmological models. Applying stability analysis and simultaneously best fitting the model with the observational data using $\chi^2$ method gives us a better understanding of the universe state. The effective EoS parameter is defined by $w_{eff}=-1-\frac{2}{3}\frac{\dot{H}}{H^{2}}$ where $\frac{\dot{H}}{H^{2}}$ is given in terms of best fitted new  dynamical variables in equation (\ref{hhdot}).

The effective EoS parameters obtained in both cases, $\gamma=0, 1/3$, do not exhibit phantom crossing behavior for different stability parameters including the best fitted ones. In case of $\gamma=0$, the EoS parameters for different model parameters including the best fitted one corresponding to the unstable critical points P1 in the past start from $w_{eff}=1$, and approaches $w_{eff}=-0.5$ in the future corresponding to the stable critical point P2 or P3. The current effective EoS parameter for different stability parameters changes and the best fitted value is about $-0.6$. Numerical solutions also reveal that
the effective EoS parameter begins to oscillate around $w_{eff}=-0.5$ at late time, as shown in the left panel of Fig. 6. The
amplitude of the oscillations are decreasing with respect to $ln (a)$.

In case of $\gamma=1/3$, the effective EoS parameters for different model parameters including the best fitted one corresponding to the unstable critical point P1 start from $w_{eff}=1$ in the past, and approach $w_{eff}=0.48$ in the future corresponding to the stable critical point P4. The current EoS parameter for different stability parameters changes and the best fitted value is $-0.79$. One of the trajectories (green curve) starts from P1 passes P2 or P3 and finally tends to infinity in the future.

\begin{tabular*}{2.5 cm}{cc}
\includegraphics[scale=.45]{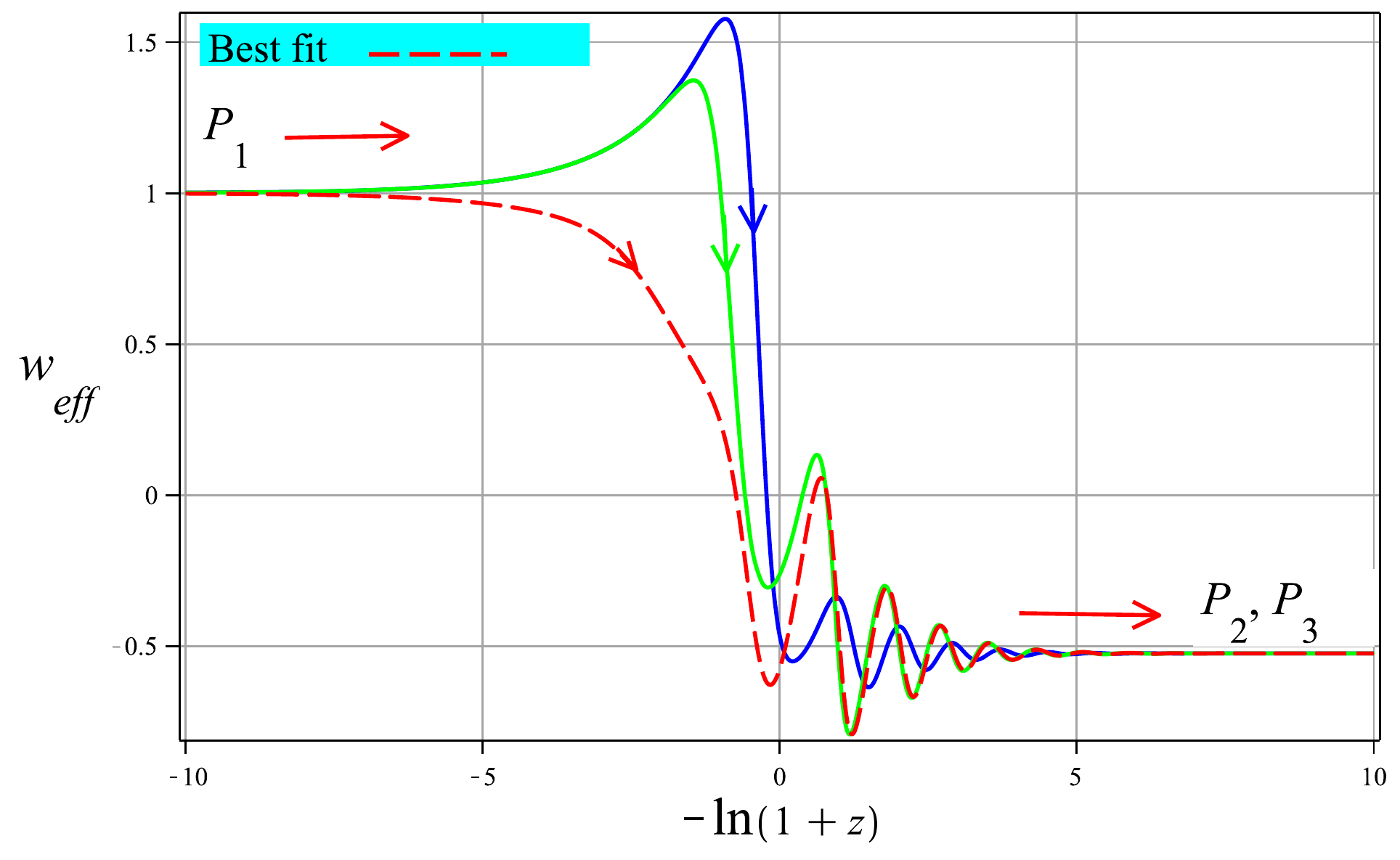}\hspace{0.1 cm}\includegraphics[scale=.45]{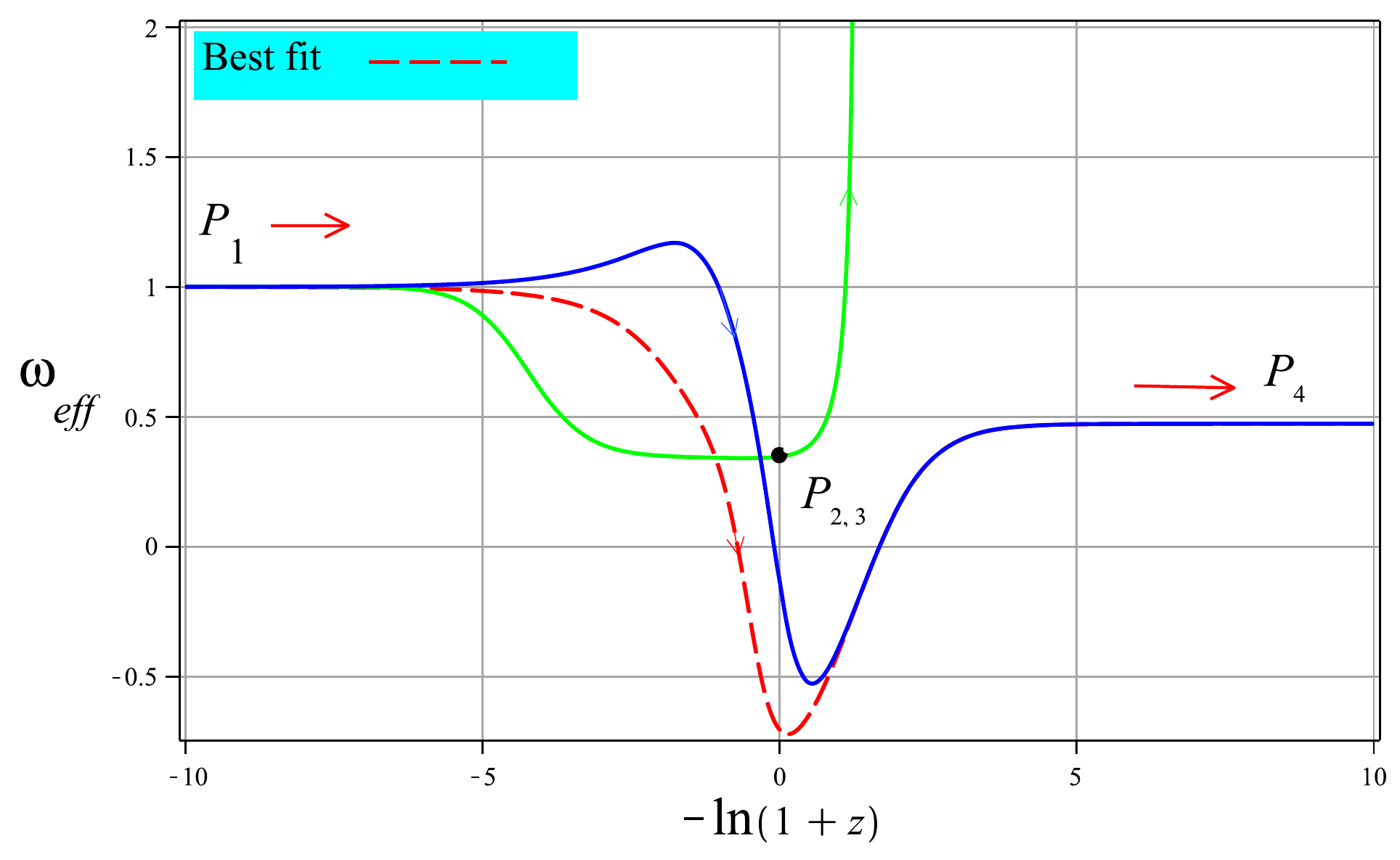}\hspace{0.1 cm}\\
Fig. 6: The effective EoS parameter $w_{eff}$, plotted as functions of redshift \\for left) $\gamma=0$ and right $\gamma=1/3$. \\
\end{tabular*}\\

\section{Discussion}

This paper is designed to study dynamics of the CGBD cosmology by
using the stability analysis and the 2-dimensional phase space of the theory. The
potential $V(\phi)$ and coupling scalar function $f(\phi)$ in the model is assumed to be in power law forms in the phase space. The matter lagrangian in the model is regarded as a perfect fluid with two kinds of EoS parameters, i.e. $\gamma=0, 1/3$. In a different approach in stability analysis, here, we solve the system of autonomous differential equations by best fitting the model parameters and also the initial conditions with the observational data for distance modulus. Therefore all the critical points with the stability conditions are presented in the model are physically reliable and observationally verified. By stability analysis, the critical points in the model for the two kind of matter fields are evaluated and shown in Figs. 5. In comparison with our previous work \cite{hossein}, in the stability analysis of CBD model where we assume a constant BD parameter, and no best fitting of the stability parameters is performed, in the current work, BD parameter is set to be an exponential function of the scalar field and best fitting of the model and stability parameters is also implemented.

We then study the cosmological parameters such as effective EoS parameter, $w_{eff}$ for the model in terms of new dimensionless dynamical variables. The model predict a late time universe acceleration expansion. While in CBD model phantom behavior occurs for the universe, in this work the best fitted effective EoS parameter does not display crossing the divide line, in both $\gamma=0, 1/3$ cases. It can be verified that the CGBD model for the constant BD parameter ($\alpha=0$) become CBD model.In comparison of the observational data for the EoS parameter of the matter, $\gamma =0, 1/3$, the results show that the current values for $w_{eff}\simeq -0.7$  corresponding to the best fitted model parameters confirm the new findings.

The oscillating behavior of the effective EoS parameter in case of $\gamma=0$, is due the spiral property of its attractor. A phase coherent or spiral attractor is one of the typical nonhyperbolic chaotic regimes that decays according to an exponential law with a decrement which is defined by the phase diffusion coefficient \cite {Vadivasova}. It has been shown numerically
and experimentally that spectral and correlation properties of these attractors can be adequately described by the model of a random
process of the harmonic noise type \cite{Vadim}. The damped oscillation for the effective EoS parameter corresponding to a universe undergoing late-time oscillating acceleration until it reaches its stable state in the future.

\end{document}